# Theoretical Analysis of Solvent Effect on NAPBr Dye's Two-photon Absorption Ability and Non-Radiative Transition in Lipid Droplets Detection

Hongyang Wang*[1], Xiaofei Wang[1], Yong Zhou[1], Jianzhong Fan*[1,2]

1. Shandong Province Key Laboratory of Medical Physics and Image Processing Technology, Institute of Materials and Clean Energy, School of Physics and Electronics, Shandong Normal University, 250014 Jinan, China
2. Guangdong Provincial Key Laboratory of Luminescence from Molecular Aggregates (South China University of Technology), 510640 Guangzhou, China

## Abstract

Two-photon fluorescence imaging has shown a promising application in biomedical imaging due to its outstanding advantages such as large penetration depth, low photo-damage, and photo-bleaching, etc. Among them, the two-photon fluorescent dye NAPBr, which can effectively select and monitor lipid droplets in living cells and biological tissues, has attracted extensive attention because of its excellent fluorescent properties. However, the research on the fluorescent abilities of two-photon fluorescent dyes in solvent environment is not sufficient. In our work, theoretical analysis reveals the internal mechanism of the solvent effect on geometric structure and photophysical properties of two-photon fluorescent dyes, especially non-radiative transition process, and holes-electrons distribution and transfer. This can provide a reference for the development of efficient two-photon absorption (TPA) molecules with aggregation-induced emission (AIE) characteristics. Related data also showed good regularity. Moreover, dye in four solvents have excellent photophysical properties: high fluorescence quantum efficiency (up to 66.60%), large Stokes shift (up to 108696 cm$^{-1}$), and two-photon absorption cross section (up to 3658 GM). The medium dielectric constant solution environment can achieve a balance between two-photon absorption and fluorescence emission capabilities better, which lays a solid foundation for the study of TPA molecules with AIE functions in terms of solvent effects.

*Author to whom correspondence should be addressed. **E-mail:** wanghongyang6787@163.com and fanjianzhongvip@163.com

# 1. Introduction

Lipid droplets (LDs) serve as an important reservoir of lipids, provide energy and substrate for membrane synthesis, whose make LDs as crucial metabolic hubs. Increasing evidence suggests that LDs are also involved in protein degradation, response to ER stress, protein glycosylation, and pathogen infection. [1-5] Recent studies have shown that lipid droplets are also highly associated with obesity, diabetes, inflammatory disorders, and cancer.[6] Therefore, the development of effective methods for selecting lipid droplets visualization and monitoring it in biological samples (such as live cells and tissues) is quiet important.[7] Compared with one-photon fluorescence imaging technology, the two-photon fluorescence imaging method utilizes two near-infrared (NIR) photons as the excitation source. It is beneficial for biomedical imaging because of deeper tissue penetration, higher spatial resolvent, lower background fluorescence, photodamage and photobleaching.[8, 9]

Tang's group used two-photon naphthalene-based donor−acceptor NAP AIEgens molecules that introduced a large heterocyclic ring to improve the electron-withdrawing ability of acceptor. The specific two-photon living cells and deep tissue lipids imaging were achieved.[7] The solvent environment affects the molecular geometric structure, photophysical property and its internal conversion. However, we have no some insight into the role of solvents in the above-mentioned Tang's research. In this work, we focus on the solvent effects on two-photon absorption and non-radiative transition. Intramolecular rotations (RIR) mainly restrict the aggregation induced emission (AIE), which affects the excited state energy consumed by non-radiative transitions, weakens NAPBr fluorescence or make it even not emits. We expound the photophysical properties, internal mechanisms with the geometrical differences caused by RIR on solvent effect. Molecules in different solvents have their unique advantages, such as excellent two-photon absorption (TPA) capability and high fluorescence quantum efficiency. It even provides a reference for the more suitable solvent environment required in actual selection. Furthermore, the solvent effect on the dye's absorption (OPA and TPA) and emission properties are studied. The molecular electron excitation is explained by analyzing electrons and holes distributions. Taking the balance between two-photon absorption and fluorescence emission ability as a standard, we find that medium dielectric constant solution is more suitable for subsequent AIE research. A comprehensive analysis of molecular property on solvent effect helps us understand the two-photon fluorescent mechanism better. It makes a solid foundation for TPA molecules research with AIE function in the solvent effect.

## 2. Theoretical calculations

In general, the oscillator strength is used to characterize the transition probability between the ground state $|i\rangle$ and the excited state $|f\rangle$ molecules in one-photon absorption and emission, which is defined as[10]

$$\delta_{op} = \frac{2\omega_f}{3}\sum_{\alpha}|\langle i|\mu_\alpha|f\rangle|^2 \qquad (1)$$

Where $\omega_f$ denotes excitation energy of the excited state $|f\rangle$, $\mu_\alpha$ is the electric dipole moment operator and the summation is performed over $x, y, z$ axes. Easy to compare with each other by defining macroscopic one-photon absorption cross sections.[11, 12]

$$\sigma_{tpa} = \frac{4\pi^2 a_0^5 \alpha}{15c} \times \frac{\omega^2 g(\omega)}{\Gamma_f}\delta_{tpa} \qquad (2)$$

Here, a0, $\alpha$, and c are the Bohr radius, the fine structure constant, and the speed of light, respectively. g(ω) denotes the spectral line profile, which is assumed to be an δ function here. Generally, the level broadening $\Gamma_f$ of the final state is assumed to be 0.1 eV, corresponding to a lifetime of about 6 fs. [13] The unit of TPA cross sections is GM (1 GM = $10^{-50}$cm$^4$ s/photon).

Moreover, the root-mean-square deviation (RMSD), which could be used to measure molecular configuration difference:

$$RMSD = \sqrt{\frac{1}{N}\sum_i^{natom}\left[(x_i - x_i^{'})^2 + (y_i - y_i^{'})^2 + (z_i - z_i^{'})^2\right]} \qquad (3)$$

N represents the number of atoms.

The fluorescence radiative decay rate can be calculated by Einstein's spontaneous emission equation:

$$K_r = \frac{\delta \Delta E_{fi}^2}{1.499 cm^{-2}} \qquad (4)$$

Here $\delta$ is the oscillator strength obtained by optimizing the first excited state (S$_1$), which is dimensionless. $\Delta E_{fi}^2$ is the vertical emission energy of S$_1$ state, the unit is cm$^{-1}$. This $K_r$ is the radiative decay rate. The unit $K_r$ is s$^{-1}$.

Based on the Fermi Golden Rule, the formula for non-radiative decay rate is

$$K_{nr} = \frac{2\pi}{\hbar^2} \sum_{u,v} P_{iv} \left| H_{fv,iv} \right|^2 \delta(E_{iv} - E_{fu}) \qquad (5)$$

$\left| H_{fv,iv} \right|$ represents the interaction between two states, $K_{nr}$ can be calculated in Molecular Materials Property Prediction Package (MOMAP) developed by Shuai group[14-18].

Based on the above-mentioned results, the fluorescent quantum efficiency can be obtained by the following equation:

$$\Phi_{PF} = \frac{K_r}{K_r + K_{nr}} \qquad (6)$$

Here, $K_r$ and $K_{nr}$ are the fluorescence rate and non-radiative decay rate of S$_1$ state.

In this article, density functional theory (DFT) are used to optimize the ground state, we use time-dependent DFT (TD-DFT) to study excited state geometry and photophysical property. Geometric configuration difference, fluorescence emission, and one-photon optical property calculations are performed based on Becke's three parametrized Lee–Yang–Parr (B3LYP) method with 6-31G* basis set. All these calculations can be achieved by a Gaussian 16 package.[19] The frequency calculations are performed to ensure optimized structure stability. The RMSD values are used to measure the molecular geometry difference between S$_0$ and S$_1$ in VMD.[20] Moreover, solution environment(such as Water, Ethanol, Acetone, and Dichloromethane) are simulated using polarizable continuum model (PCM).[21] Corresponding dielectric constant (ε) of four solvents are listed in Table 1,which can be found in the Gaussian manual.

Table 1. The dielectric constant (ε) of four solvents.

| Solvent | Water | Ethanol | Acetone | Dichloromethane |
|---|---|---|---|---|
| ε | 78.36 | 32.61 | 20.49 | 8.93 |

Furthermore, equilibrium solvation is applied for geometry optimization because solvent needs time to fully respond to the changes of solute in two ways: polarizing its electronic distribution and making its nuclei reoriented.[22] Under the frame of first-order perturbation theory, non-adiabatic electronic coupling is treated as the force acting on atomic nuclei through transition electric field, which is evaluated in this paper.[22,23] The electron distribution and frontier molecular orbital (FMO) levels are analyzed. Besides, molecular fluorescent quantum efficiency, Huang−Rhys (HR) factor, and reorganization energy in different solvents are all calculated by MOMAP based on organic fluorescent molecules excited state decay theory. The two-photon absorption (TPA) properties are calculated by Datlon2013.[24]

3. Result and discussion

3.1 Molecular structures

Firstly, we choose NAPBr in the experiment performed by Tang's group and study molecular structures in Water, Ethanol, Acetone, and Dichloromethane.[7] Br is choosing for its strong tendency to get electrons and it is one of the most electronegative elements, which is more convenient for later analysis. The structural formula of studied compound, optimized ground state structure is shown in Figure 1.

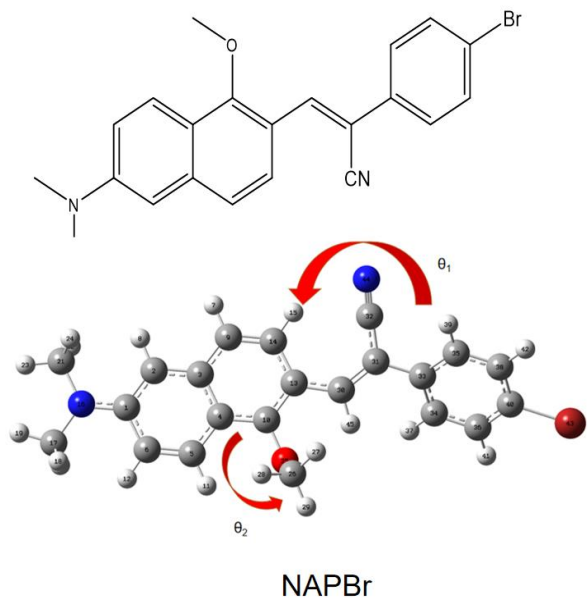

Figure 1 Structural formula of the studied compound and its geometric structure. Atomic labels and interesting angles are marked. Dihedral angle $\theta_1$: $C_{14}$-$C_{13}$-$C_{33}$-$C_{35}$, bond angle $\theta_2$: $C_{10}$-$C_{25}$-$C_{26}$.

In this part, we study the solvent effect on molecular geometrical structures. Dihedral angle $\theta_1$ between the benzene ring and naphthalene, bond angle $\theta_2$ are selected for analysis. With decreasing solvent polarity, it can be found that the $\theta_1$ are 53.30°, 53.26°, 53.20°, 53.07°, $\theta_2$ are 114.99°, 114.97°, 114.96°, 114.91°. Dihedral angle of NAPBr in Water is the largest, while that in Dichloromethane is the smallest. Larger solvent environment polarity causes a smaller twist angle difference between naphthalene and benzene ring, molecular co-planar structure is destroyed slightly. Therefore, molecular electron distribution change lead to a discrepancy in optical properties.

The OPA and OPE of dyes

Absorption and emission properties are important nature for dyes. In this chapter, we study the OPA and OPE properties and transition process of NAPBr in different solvents in detail.

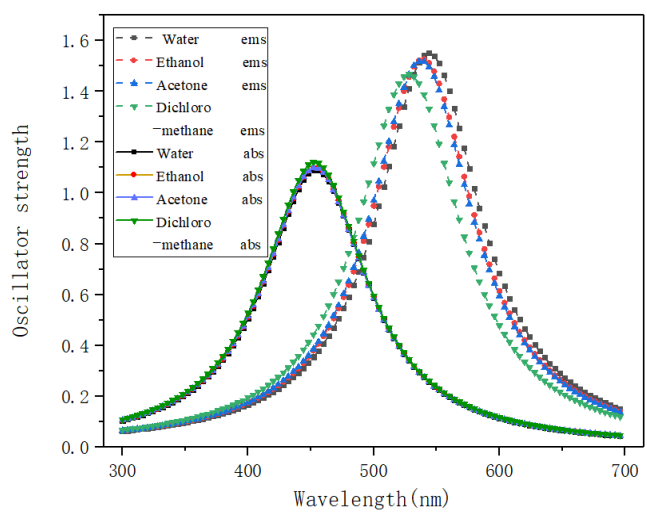

Figure 2 The relationship between absorption, emission wavelength, and oscillator strength in different solvents.

Table 2 Detailed data on molecular one-photon absorption and fluorescent emission, a: emission wavelength maximum, b: absorption wavelength maximum.

| Molecule | $\theta_1$ | $\theta_2$ | $E_{flu}$ | $\lambda^a_{ems}$ | $\delta_{ems}$ | $\lambda^b_{abs}$ | $\delta_{abs}$ | Stokes shift | Transition nature |
|---|---|---|---|---|---|---|---|---|---|
| Water | 53.30° | 114.99° | 2.28 | 544 | 1.55 | 454 | 1.09 | 90 nm | S1→S0 H→L 99% |
| Ethanol | 53.26° | 114.97° | 2.30 | 539 | 1.53 | 454 | 1.10 | 85nm | S1→S0 H→L 99% |
| Acetone | 53.20° | 114.96° | 2.31 | 538 | 1.52 | 453 | 1.10 | 85 nm | S1→S0 H→L 99% |
| Dichloromethane | 53.07° | 114.91° | 2.34 | 528 | 1.47 | 452 | 1.12 | 75 nm | S1→S0 H→L 99% |

Then, molecular OPA and OPE properties are studied and compared in four solvents. As solvent polarity decreases, it is easy to find that molecular emission wavelengths are 544 nm, 539 nm, 538 nm, 528 nm, after analyzing Figure 3 Molecular OPA process. The absorption wavelengths are 454 nm, 454 nm, 453 nm, 452 nm, which is gradual decreases. NAPBr in Water has maximum emission and absorption wavelength. Specific data are collected in Table . In the experimental data, emission wavelength and absorption wavelength are 525 nm 409 nm, they are within reasonable limits with calculation. Different excitation wavelengths, interaction between laser and molecules, or inter-molecular interaction (such as solute-solute interaction, etc.) in solution, could result in the discrepancy and have not been considered in the calculation.[25, 26] Molecular absorption and emission wavelengths in other solvents show blueshift to that in Water. They have long Stokes shifts (the difference between the absorption and emission maximum wavelength) over 75nm, up to 90 nm. It has been known that long Stokes shift helps to avoid the interference of absorption and emission spectra and improves detection sensitivity and accuracy.[27] It is related to the maximum energy gap of HUMO-LUMO (HOMO represents the highest occupied molecular orbital and LUMO represents the lowest unoccupied molecular orbital). $E_{flu}$ is the energy difference between the two electronic states. $\delta_{abs}$ and $\delta_{ems}$ are absorption and emission oscillator strength which are important physical parameters that measure absorption or emission ability. Intramolecular charges transfer is less affected by solvent, so the energy gap between two electronic states changes slightly. Besides, electrical transition dipole moment is also important. Figure 2 shows that solvent polarity has a great influence on molecular fluorescent emission properties. But it has less influence on the variation of absorption wavelength and oscillator strength.

Now, we perform a detailed analysis of OPA and OPE processes and their transition properties to show molecular optical properties visually.[28] The OPA (or OPE) wavelength, corresponding oscillator strength, molecular orbital energy, and transition natures have been given.

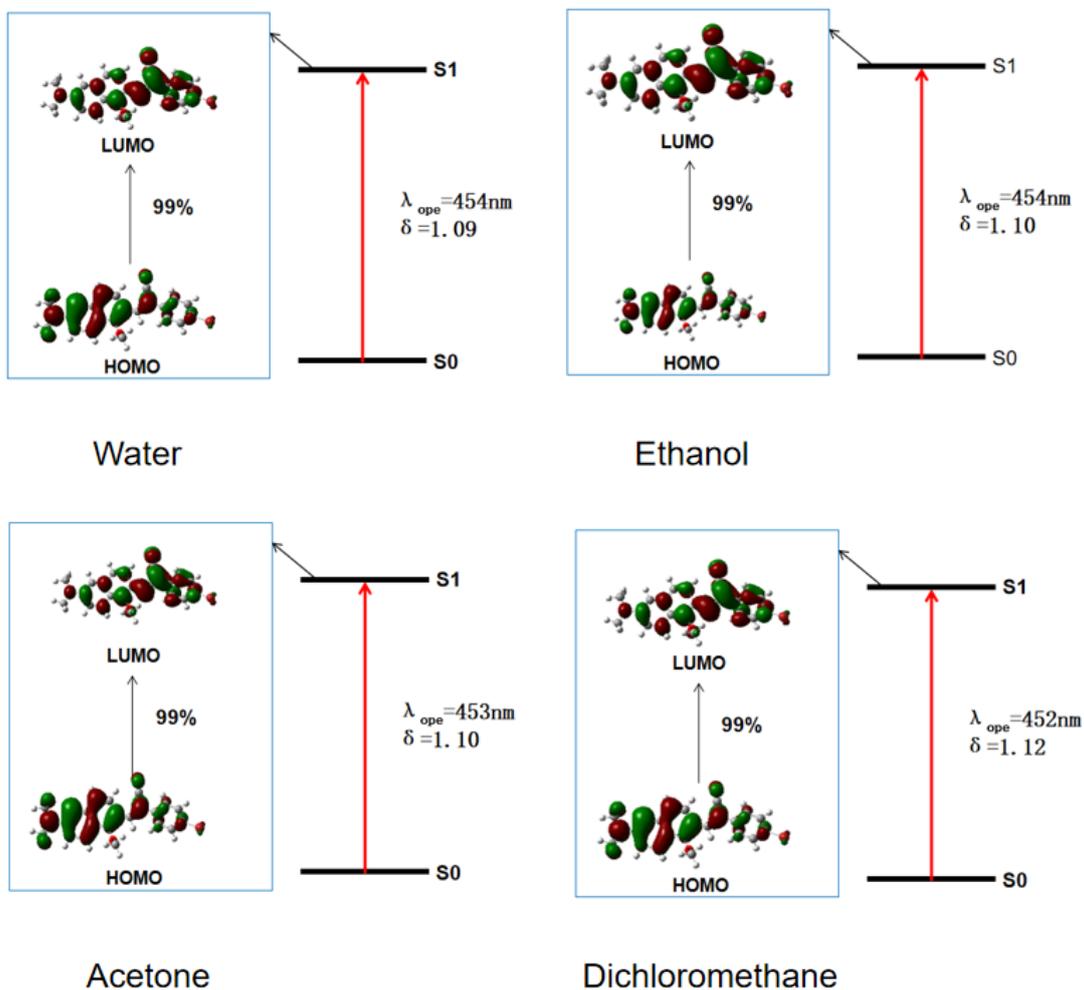

Figure 3 Molecular OPA processes.

First, we analyze the molecular absorption process in detail. Red and green represent positive and negative phases of molecular orbitals. Through studying transition process, the maximum OPA state is $S_1$. 99% of molecular orbitals are originated from the HOMO to LUMO transition. It is easy to find that molecular orbitals changes are mainly localized on the torsional benzene ring moiety in Figure 3, which is closely connected with the absorption process.

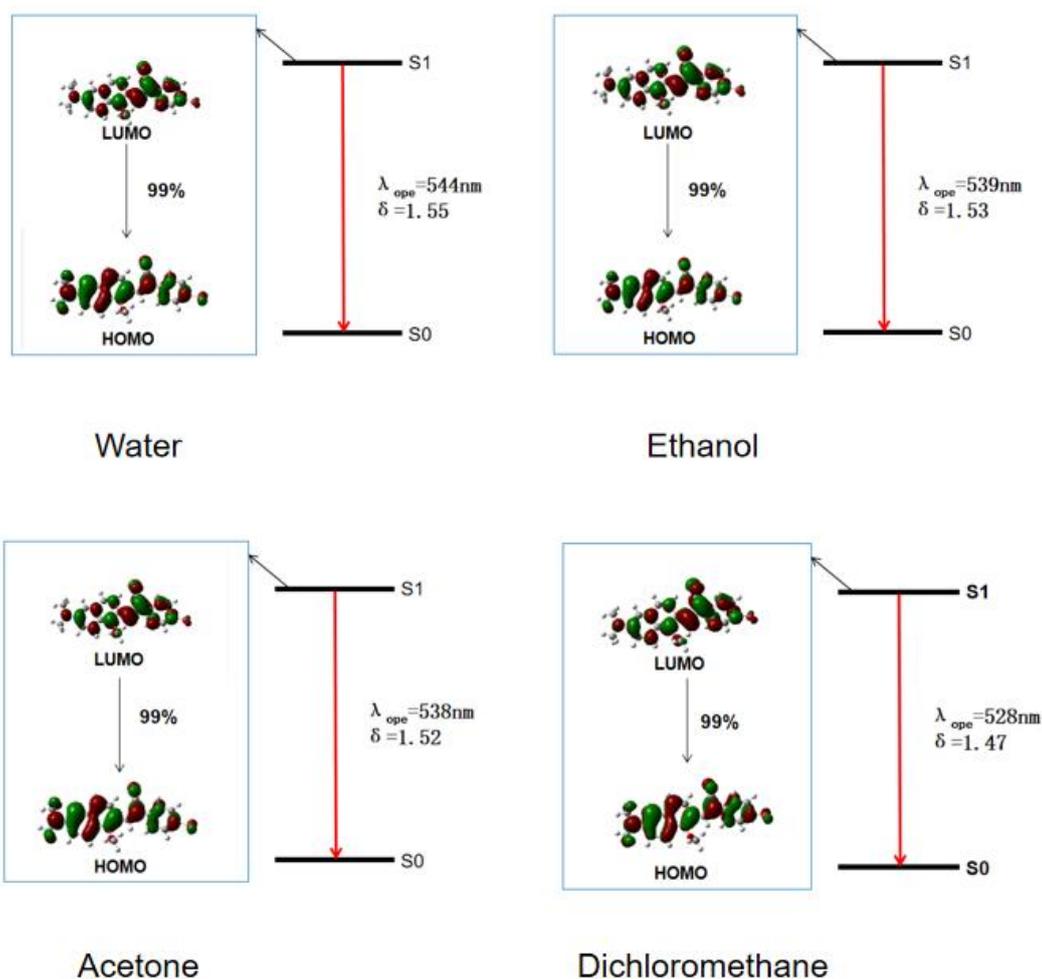

Figure 4  molecular OPE processes.

In Figure 4, the OPE processes are shown. According to Kasha's rule, the fast and non-radiative internal conversion and vibrational relaxation process result in being located at the lowest molecular vibrational level in $S_1$ after excitation, which also affects the molecular configuration. It can be seen that the $S_1$ to $S_0$ transition mostly results from the LUMO to HOMO transition. Molecular orbital distribution is similar in the twisted benzene ring and mainly located on the naphthol, which is fluorescent group. Small torsional benzene ring orbitals variation indicate that this part has a little effect on the excitation. OPE processes under different solvents are similar.

Holes and electrons analysis

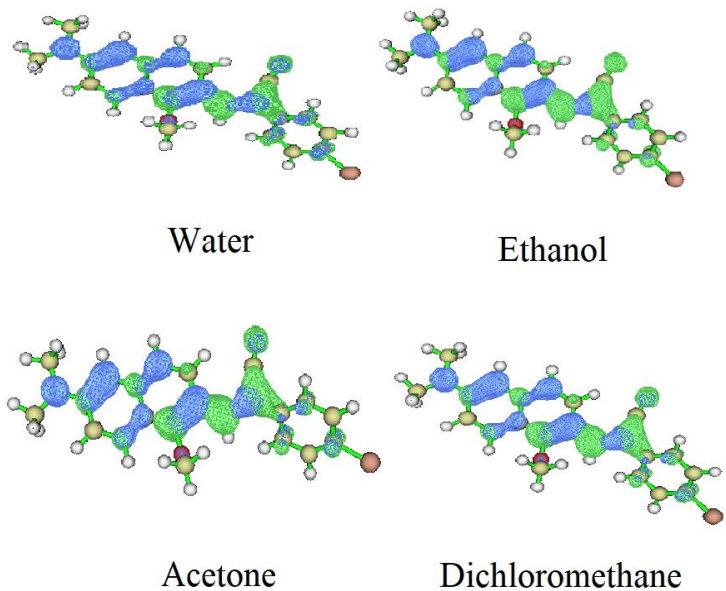

Figure 5 Molecular holes and electrons distribution

Table 3 Detailed data of molecular holes and electrons distribution, including D index, $S_r$ and $\Delta\sigma$

| Solvent | D index/Å | $S_r$/. a. u | $\Delta\sigma$/ Å |
|---|---|---|---|
| Water | 1.853 | 0.77719 | -0.401 |
| Ethanol | 1.854 | 0.77720 | -0.402 |
| Acetone | 1.855 | 0.77719 | -0.402 |
| Dichloromethane | 1.857 | 0.77723 | -0.404 |

Define D index that measures the distance between hole and electron centroid:

$$D_x = | X_{ele} - X_{hole} |$$

$$D_y = | Y_{ele} - Y_{hole} |$$

$$D_z = | Z_{ele} - Z_{hole} |$$

$$Dindex = \sqrt{(D_x)^2 + (D_y)^2 + (D_y)^2}$$

$X_{ele/hole}$ refers to the X coordinate of the mass center of hole or electron.

Define $S_r$ as a function that describes the distribution overlap between electrons and holes:

$$S_r(r) = \sqrt{\rho^{hole}(r)\rho^{ele}(r)}$$

$\Delta\sigma$ reflects the difference in the overall spatial distribution of electrons and holes:

$$\Delta\sigma index = |\sigma_{ele}| - |\sigma_{hol}|$$

We draw molecular holes and electrons distribution diagram to study electrons excitation characteristics. Green and blue represent the distribution of electrons and holes, which are almost distributed in the naphthalene plane with no obvious difference. Electrons converge near the torsional benzene ring, while holes are away. Detailed data are listed in Table 3. Large D index means large distance between electrons and holes. Large $S_r$ and small $\Delta\sigma$ indicate large overlap between holes and electrons. Electrons - holes separation is not sufficient, and overall spatial distribution discrepancy is small. This is considered to be a charge transfer excitation. Besides, the main excitation process is from $S_0$ to $S_1$. NAPBr in Dichloromethane has the smallest $\Delta\sigma$, largest D index and $S_r$. It is closely related to non-radiative transition process, which may explain the reason that has the largest fluorescence quantum efficiency. Dye's fluorescent property is structure-dependent, especially for the donor−acceptor molecule. It is also confirmed that donor and acceptor transfer charges to achieve the fluorescence.

Two-photon absorption property

Excellent molecular TPA ability is an important criterion for high-quality fluorescent dyes. Therefore, we calculate the NAPBr TPA property (such as excitation energy, corresponding TPA cross section, and wavelength) in different solvents, as shown in

Table 4 Two-photon absorption properties including excitation energy Etpa, corresponding two-photon wavelength λtpa, and TPA cross section σtpa (1 GM = 10-50cm4 s/photon) of the lowest five excited states of dye in different solvents.. Lowest five molecular excited states are selected for analysis.

Table 4 Two-photon absorption properties including excitation energy $E_{tpa}$, corresponding two-photon wavelength $\lambda_{tpa}$, and TPA cross section $\sigma_{tpa}$ (1 GM = $10^{-50}$cm$^4$ s/photon) of the lowest five excited states of dye in different solvents.

| Solvent | $E_{tpa}$/ eV | $\lambda_{tpa}$/nm | $\sigma_{tpa}$/GM | Solvent | $E_{tpa}$/ eV | $\lambda_{tpa}$/nm | $\sigma_{tpa}$/GM |
|---|---|---|---|---|---|---|---|
| Water | 2.61 | 948 | 519 | Ethanol | 2.63 | 940 | 478 |
|  | 3.41 | 725 | 10 |  | 3.42 | 723 | 10 |
|  | 3.68 | 672 | 18 |  | 3.69 | 670 | 4 |
|  | 3.76 | 658 | 3658 |  | 3.76 | 658 | 3461 |
|  | 4.10 | 603 | 9 |  | 4.11 | 602 | 6 |
| Acetone | 2.64 | 937 | 481 | Dichloromethane | 2.66 | 930 | 435 |
|  | 3.42 | 723 | 10 |  | 3.42 | 723 | 9 |
|  | 3.7 | 668 | 7 |  | 3.71 | 667 | 4 |
|  | 3.77 | 656 | 3375 |  | 3.78 | 654 | 3016 |
|  | 4.11 | 602 | 6 |  | 4.11 | 602 | 4 |

As solvent polarity increases, the increases maximum molecular TPA cross section in different solvents are 3016GM (Dichloromethane, 654nm), 3375GM (Acetone, 656nm), 3461GM (Ethanol, 658nm), 3658GM (Water, 658nm). Maximum molecular TPA cross section is located at the lowest fourth excited state, regardless of solvent environment. It indicates that solvent does not affect TPA cross section distribution and there has a certain stability. Compared with molecules in other solvents, it can be seen that NAPBr in Water has a longer OPA wavelength, larger TPA cross section, and OPA oscillator strength. Otherwise, a good conjugated planarity system and strong acceptor group are important for excellent molecular absorption ability in Tang's group analysis.[10] In addition, we calculate the electric transition dipole moments from $S_0$ and $S_1$ are 3. 8221.a. u (Water), 3. 8212. a. u (Ethanol), 3. 8204. a. u (Acetone), 3. 8177. a. u (Dichloromethane), they are closely related to molecular absorption process. Besides, electric transition dipole moment between $S_0$ and $S_1$ is proportional to the orbital overlap of HOMO and LUMO. TPA ability, OPA oscillator strength orders are consistent with electric transition dipole moment, which have regular and certain stability in solvent. It can be expected that NAPBr can act as a promising candidate for a two-photon fluorescent dye.

Fluorescent quantum efficiency of dyes

Table 5 Fluorescence properties, including vertical excitation energy ($E_{vt}$), adiabatic energy difference between S0 and S1 ($E_{ad}$), radiative decay rate ($k_r$), non-radiative decay rate ($k_{ic}$), RMSD, reorganization energy ($E_{re}$) and fluorescence quantum efficiency (Φ) of molecules in different solvents.

| Solvent | $E_{vt}$/eV | $E_{ad}$/eV | $k_r$/s$^{-1}$ | $k_{ic}$/s$^{-1}$ | $\Phi$ | $E_{re}$/cm$^{-1}$ | RMSD/Å |
|---|---|---|---|---|---|---|---|
| Water | 2.28 | 2.42 | 3.498× 10$^8$ | 2.278× 10$^8$ | 60.56% | 4987.02 | 0.23 |
| Ethanol | 2.30 | 2.44 | 3.506× 10$^8$ | 2.08× 10$^8$ | 62.74% | 4813.63 | 0.228 |
| Acetone | 2.31 | 2.45 | 3.508× 10$^8$ | 2.018× 10$^8$ | 63.48% | 4754.04 | 0.228 |
| Dichloromethane | 2.35 | 2.48 | 3.509× 10$^8$ | 1.759× 10$^8$ | 66.60% | 4399.03 | 0.224 |

Fluorescent quantum efficiency is also an important index for evaluating dyes properties. According to Kasha's law, the radiation competition from $S_1$ to $S_0$ has an important impact on fluorescence quantum efficiency as well as adiabatic excitation energy. From Table , it is easy to find that as solvent polarity gradually decreases, the fluorescent quantum efficiency order is 60.56% (Water) ＜62.74%( Ethanol ) ＜63.48% (Acetone) ＜66.60% (Dichloromethane). Radiative decay rates are closely related to the solvent polarity, they are 3.498× 10$^8$s$^{-1}$＜3.506× 10$^8$s$^{-1}$＜3.508× 10$^8$s$^{-1}$＜3.509× 10$^8$s$^{-1}$, respectively. The calculated NAPBr fluorescence quantum efficiency is 67.6% and the experimental result is 1.4%. AIE, solvent, and solute-solute interaction are important factors.[7] The NAPBr in Dichloromethane has the highest fluorescent quantum efficiency because it's highest radiative decay rate (3.509× 10$^8$s$^{-1}$) and lowest non-radiative decay rate (1.759 × 1 0$^8$ s$^{-1}$). Molecules in low dielectric constant solvent environment can suppress the non-radiative loss path of excited-state energy, thereby improving fluorescent quantum efficiency. [28] Besides, electrons and holes distribution and transfer analysis also help us explain the reason that NAPBr has a higher fluorescence quantum efficiency in Dichloromethane. After considering all aspects of performance, we think that NAPBr has a higher fluorescent quantum efficiency, larger TPA cross-section, and Stokes shift. For the underlying reasons of solvent effect on fluorescent quantum efficiency, we conduct a detailed study in the next section.

Reorganization energy and Huang-Rhys factor

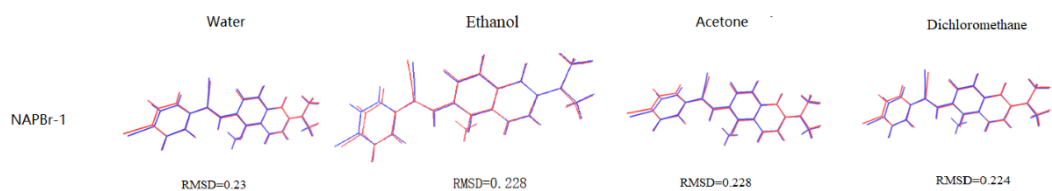

Figure 6 RMSD between the two states configuration in Water, Ethanol, Acetone, and Dichloromethane. Blue represents $S_0$ configuration and red represents $S_1$ configuration.

This chapter studies the solvent effect on non-radiative internal conversion from $S_0$ to $S_1$ in detail. This is the key to explain the inner reason of fluorescence quantum efficiency. Firstly, RMSD is calculated and compared to illustrate the solvent effect on molecular configuration difference between $S_0$ and $S_1$ clearly as shown in Figure 6. As solvent polarity decreases, the RMSD are 0.23 Å (Water), 0.228 Å (Ethanol), 0.228 Å (Acetone), 0.224 Å (Dichloromethane). Molecular geometric changes mostly originate from benzene ring rotation, such as length and dihedral angle change. They are closely related to non-radiative transition process and can directly prove the necessity of studying the dihedral angle $\theta_1$.

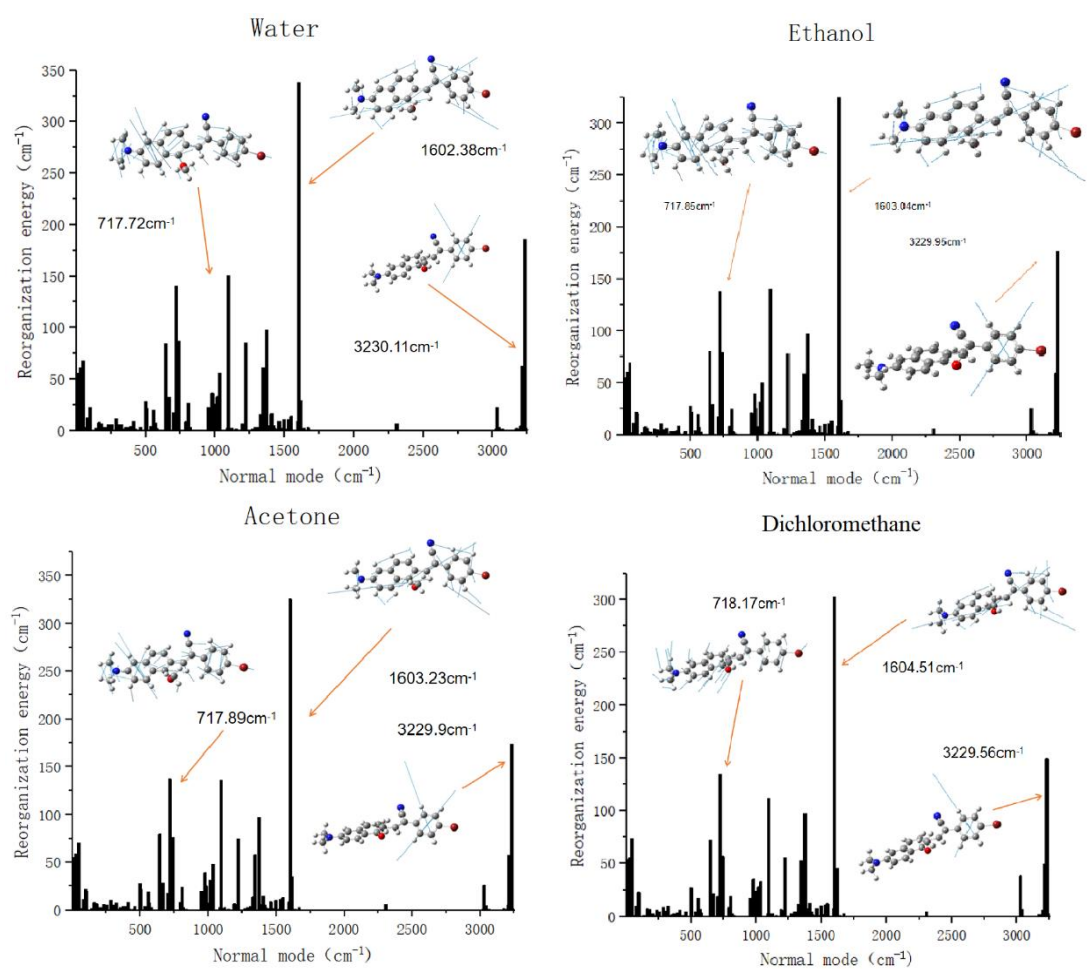

Figure 7  Molecular reorganization energy distribution under different vibration modes in different solvents.

Non-radiative decay rate is related to the geometric difference between two states, also affects fluorescent quantum efficiency. The non-radiative decay rates of NAPBr are $2.28\times10^8 s^{-1}$(Water), $2.08\times10^8 s^{-1}$(Ethanol), $2.02\times10^8 s^{-1}$(Acetone) and $1.76\times10^8 s^{-1}$(Dichloromethane). Non-radiative decay rate is more sensitive to solvent polarity than the radiative decay rate. NAPBr in Water has maximum non-radiative decay rate ($2.278\times 10^8 s^{-1}$) and minimum radiative decay rate ($3.498\times 10^8\ s^{-1}$). So, it has the lowest fluorescent quantum efficiency. Geometric relaxation degree between two states and internal conversion energy attenuation can be characterized by reorganization energy, as shown in Figure 7. Obviously, it is easy to find that the reorganization energy contributions of low-frequency modes region dominate vibrational relaxation processes, which is easily affected by the solvent polarity. As the solvent polarity decreases, the reorganization energies are 4987.02 cm$^{-1}$ (Water), 4813.63 cm$^{-1}$ (Ethanol), 4753.04 cm$^{-1}$ (Acetone), and 4399.03 cm$^{-1}$ (Dichloromethane), receptively. Reorganization energy peak value in high-frequency modes region also decreases. Larger reorganization energy in low-frequency modes region may be attributed to structure changes, especially benzene ring rotation. Molecular difference is mainly reflected in the dihedral angle $\theta_1$ caused by solvent effect. It is closely related to the geometric structure difference of $S_0$ and $S_1$ in the non-radiative transition.

According to Marcus Theory, if no bond is formed or broken in electrons transfer reaction, it has a less geometric relaxation, lower reorganization energy, and higher electron transfer rate. The consequence of electrons transfer is charges rearrangement and it is greatly influenced by the solvent environment. Moreover, a D-π-A structure molecule provide a very important factor for charges transfer. It is the solvent polarity that determines free activation energy in internal conversion, thereby affecting molecular photophysical properties. Molecule with different geometric configurations in two states transforms firstly into a more similar shape by vibration and then transfer electrons to make the geometry equivalent. It can also be used to explain the rule of fluorescent quantum efficiency. For example, the NAPBr in Dichloromethane has less difference in $S_0$ and $S_1$ geometries, so it can transfer electrons with a little vibration.

Finally, we study Huang−Rhys (HR) factors to measure tnon-radiative excited state energy consumption, characterize electron vibration coupling, and the contribution degree of geometric configuration difference at this frequency. It is helpful for further analyzing the vibration situation and the non-radiative transition comprehensively and holistically. Some vibration modes with larger HR factor (> 0.1) have been listed.[29] The Huang−Rhys (HR) factor:
$$HR_K = \frac{\omega_K D_K^2}{2}$$

$\omega_k$ represents the vibration frequency and $D_k$ is the normal coordinate displacement of mode k.

Table 6 Relatively bigger Huang−Rhys (HR) factor (＞0.1) and reorganization energy in some vibration mode and frequency.

| Solvent | Mode | f | HR | $E_{re}/cm^{-1}$ | Solvent | Mode | f | HR | $E_{re}/cm^{-1}$ |
|---|---|---|---|---|---|---|---|---|---|
| Acetone | 1 | 16.39 | 3.35 | 54.93 | Dichlorom-ethane | 1 | 16.53 | 3.26 | 53.86 |
| | 2 | 22.03 | 0.18 | 3.90 | | 2 | 22.5 | 0.44 | 9.81 |
| | 3 | 29.95 | 1.97 | 58.97 | | 3 | 30.1 | 1.84 | 55.41 |
| | 4 | 39.97 | 0.17 | 6.98 | | 4 | 39.62 | 0.16 | 6.43 |
| | 5 | 50.15 | 1.39 | 69.87 | | 5 | 50.72 | 1.46 | 73.89 |
| | 8 | 81.39 | 0.14 | 11.11 | | 8 | 81.71 | 0.11 | 9.39 |
| | 9 | 96.5 | 0.23 | 22.02 | | 9 | 96.67 | 0.23 | 22.02 |
| | 10 | 100.83 | 0.20 | 20.29 | | 10 | 101.17 | 0.22 | 22.73 |
| | 39 | 642.73 | 0.12 | 79.59 | | 39 | 642.98 | 0.11 | 71.81 |
| | 44 | 717.89 | 0.19 | 137.20 | | 44 | 718.17 | 0.19 | 133.86 |
| | 67 | 1094.41 | 0.12 | 136.30 | | 67 | 1094.88 | 0.10 | 111.78 |
| | 105 | 1603.23 | 0.20 | 325.23 | | 105 | 1604.51 | 0.19 | 302.51 |

| Solvent | Mode | f | HR | $E_{re}/cm^{-1}$ | Solvent | Mode | f | HR | $E_{re}/cm^{-1}$ |
|---|---|---|---|---|---|---|---|---|---|
| Water | 1 | 16.27 | 3.43 | 55.81 | Ethanol | 1 | 16.37 | 3.37 | 55.09 |
| | 3 | 29.88 | 2.04 | 61.02 | | 2 | 21.94 | 0.14 | 3.12 |
| | 4 | 40.72 | 0.19 | 7.78 | | 3 | 29.94 | 1.99 | 59.64 |
| | 5 | 49.9 | 1.35 | 67.28 | | 4 | 40.11 | 0.18 | 7.16 |
| | 8 | 81.17 | 0.15 | 12.05 | | 5 | 50.09 | 1.38 | 69.16 |
| | 9 | 96.43 | 0.23 | 22.40 | | 8 | 81.34 | 0.14 | 11.24 |
| | 10 | 100.56 | 0.18 | 18.53 | | 9 | 96.48 | 0.23 | 22.14 |
| | 39 | 642.58 | 0.13 | 84.35 | | 10 | 100.77 | 0.20 | 19.85 |
| | 44 | 717.72 | 0.20 | 140.07 | | 39 | 642.70 | 0.13 | 80.76 |
| | 46 | 739.45 | 0.12 | 86.62 | | 44 | 717.85 | 0.19 | 137.88 |
| | 67 | 1094.12 | 0.14 | 150.01 | | 67 | 1094.34 | 0.13 | 140.03 |
| | 105 | 1602.38 | 0.21 | 337.65 | | 105 | 1603.04 | 0.20 | 327.89 |

Larger HR factors are distributed in the low-frequency mode (<500 cm$^{-1}$) region and have different contribution degree of structural changes, this is present in all four solvents. [28] The largest HR factors are 3.43 (16.27 cm$^{-1}$, Water), 3.37 (16.37 cm$^{-1}$, Ethanol), 3.35 (16.39 cm$^{-1}$, Acetone), 3.26 (16.53 cm$^{-1}$, Dichloromethane). Vibration frequency and its corresponding HR factor change slightly. Total HF factor reduces and shows that molecular rotational motion is suppressed. It indicates that solvent polarity has a slight effect on the HR factor and its frequency mode. Geometric configuration difference, non-radiative transition, and total HR factor are closely related and well characterized. This also proves that solvent polarity affects the molecular photophysical performance, further confirms our calculation result, and clarifies research's necessity.

Conclusion

In this work, we conduct a theoretical study on the solvent effect of naphthalene-based dyes for detecting lipid droplets at the DFT level. Result shows that NAPBr exhibits excellent and regular two-photon fluorescence ability in all four solvents, with a high fluorescence quantum efficiency, large Stokes shift, and TPA cross-section. The internal mechanism of solvent effect on photophysical properties is also analyzed in detail. More than 99% molecular orbitals originate from the HOMO—LUMO transition in OPA and OPE process. Electrons and holes analysis prove that fluorescence emission mechanism is a charge transfer process between donor and acceptor. Through affecting non-radiative transition process, molecular structure variation (especially benzene ring rotation) degree in different solvents is closely related to the fluorescence quantum efficiency. After considering all aspects of performance, a medium dielectric constant solvent environment can achieve a balance of two-photon absorption and fluorescence emission ability. The conclusion and method are more suitable to provide a reference and help for the subsequent TPA dyes research with AIE characteristics in the future.

Reference

1. Hartman, I. Z.; Liu, P.; Zehmer, J. K.; Luby-Phelps, K.; Jo, Y.; Anderson, R. G.; DeBose-Boyd, R. A. *J Biol Chem* **2010,** 285, (25), 19288-98.
2. Fei, W.; Wang, H.; Fu, X.; Bielby, C.; Yang, H. *Biochem J* **2009,** 424, (1), 61-7.
3. Olzmann, J. A.; Richter, C. M.; Kopito, R. R. *Proc Natl Acad Sci U S A* **2013,** 110, (4), 1345-50.
4. Krahmer, N.; Hilger, M.; Kory, N.; Wilfling, F.; Stoehr, G.; Mann, M.; Farese, R. V., Jr.; Walther, T. C. *Mol Cell Proteomics* **2013,** 12, (5), 1115-26.
5. Herker, E.; Ott, M. *Trends Endocrinol Metab* **2011,** 22, (6), 241-8.
6. Krahmer, N.; Farese, R. V.; Walther, T. C. J. E. m. m. **2013,** 5, (7), 973-983.
7. Niu, G.; Zhang, R.; Kwong, J. P.; Lam, J. W.; Chen, C.; Wang, J.; Chen, Y.; Feng, X.; Kwok, R. T.; Sung, H. H.-Y. J. C. o. M. **2018,** 30, (14), 4778-4787.
8. Kim, H. M.; Cho, B. R. *Chemical Reviews* **2015,** 115, (11), 5014-5055.
9. Pawlicki, M.; Collins, H. A.; Denning, R. G.; Anderson, H. L. *Angewandte Chemie International Edition* **2009,** 48, (18), 3244-3266.
10. Hilborn, R. C. J. A. J. o. P. **1982,** 50, (11), 982-986.
11. Luo, Y.; Norman, P.; Macak, P.; Ågren, H. J. T. J. o. P. C. A. **2000,** 104, (20), 4718-4722.
12. Wang, C.-K.; Zhao, K.; Su, Y.; Ren, Y.; Zhao, X.; Luo, Y. J. T. J. o. c. p. **2003,** 119, (2), 1208-

1213.

13. Albota, M.; Beljonne, D.; Brédas, J.-L.; Ehrlich, J. E.; Fu, J.-Y.; Heikal, A. A.; Hess, S. E.; Kogej, T.; Levin, M. D.; Marder, S. R. J. S. **1998,** 281, (5383), 1653-1656.
14. Niu, Y.; Li, W.; Peng, Q.; Geng, H.; Yi, Y.; Wang, L.; Nan, G.; Wang, D.; Shuai, Z. J. M. P. **2018,** 116, (7-8), 1078-1090.
15. Niu, Y.; Peng, Q.; Shuai, Z. J. S. i. C. S. B. C. **2008,** 51, (12), 1153-1158.
16. Peng, Q.; Yi, Y.; Shuai, Z.; Shao, J. J. J. o. t. A. C. S. **2007,** 129, (30), 9333-9339.
17. Shuai, Z.; Peng, Q. J. N. S. R. **2017,** 4, (2), 224-239.
18. Shuai, Z.; Peng, Q. J. P. R. **2014,** 537, (4), 123-156.
19. GAUSSIAN 16, R. i. h. w. g. c. GAUSSIAN 16, References in http://www.gaussian.com/
20. Humphrey, W.; Dalke, A.; Schulten, K. J. J. o. m. g. **1996,** 14, (1), 33-38.
21. Tomasi, J.; Mennucci, B.; Cammi, R. J. C. r. **2005,** 105, (8), 2999-3094.
22. Fan, D.; Yi, Y.; Li, Z.; Liu, W.; Peng, Q.; Shuai, Z. *The journal of physical chemistry. A* **2015,** 119, (21), 5233-40.
23. Lin, S. H. *The Journal of Chemical Physics* **1966,** 44, (10), 3759-3767.
24. DALTON 2013 A molecular electronic structure program, R. D., see http://daltonprogram.org/.
25. Zhang, Y.; Leng, J.; Hu, W. *Sensors* **2018,** 18, (5).
26. Zhang, Y.-J.; Wang, X.; Zhou, Y.; Wang, C.-K. *Chemical Physics Letters* **2016,** 658, 125-129.
27. Zhu, M.-Y.; Zhao, K.; Song, J.; Wang, C.-K. J. C. P. B. **2018,** 27, (2), 023302.
28. Fan, J.; Zhang, Y.; Zhou, Y.; Lin, L.; Wang, C.-K. J. T. J. o. P. C. C. **2018,** 122, (4), 2358-2366.
29. Cai, W.; Zhang, H.; Yan, X.; Zhao, A.; He, R.; Li, M.; Meng, Q.; Shen, W. J. P. C. C. P. **2019,** 21, (15), 8073-8080.